\documentclass[aps,prl,showpacs,floats,twocolumn,amsfonts,amssymb,unsortedaddress,floatfix]{revtex4}
\input epsf
\usepackage{bm}
\usepackage{amsmath}
\usepackage{graphicx}

\begin{document}

\title{Thermally Induced Local Failures in Quasi-One-Dimensional Systems: Collapse in  Carbon Nanotubes, Necking in Nanowires and Opening of Bubbles in DNA. }
	
\author{Cristiano~Nisoli$^{1}$, Douglas Abraham$^{1,2}$, Turab Lookman$^{1}$ and Avadh Saxena$^{1}$}
\affiliation{$^{1}$\mbox{Theoretical Division and Center for Nonlinear Studies, Los Alamos National Lab, Los Alamos NM 87545 USA} \\
$^{2}$Rudolf Peierls Centre for Theoretical Physics, 1 Keble Road Oxford, OX1 3NP England }

\date{\today}
\begin{abstract}
We present a general framework to explore thermally activated failures in quasi one dimensional systems. We apply it to the collapse of carbon nanotubes, the formation of  bottlenecks in nanowires, both of which limit conductance, 
and the opening of local regions or "bubbles" of base pairs in strands of DNA that are relevant for transcription and danaturation.  We predict an exponential behavior for the probability of the opening
 of bubbles in DNA, the average distance between flattened regions of a  nanotube or necking in a nanowire as a monotonically decreasing function of temperature, and  compute a temperature below which these events become extremely rare. These findings are difficult to obtain numerically, however, they could be accessible experimentally.  
  
\end{abstract}


\maketitle

Nanowires \cite{wires}, carbon  nanotubes~\cite{ji} and DNA filaments~\cite{prot} are examples of Quasi One Dimensional (Q1D) mesoscopic systems characterized by one very long dimension, on the molecular length scale, compared to the others that are of the order of nanometers. Their  theoretical and technological interest is very high and motivates an analysis of the mechanisms governing their failure, yet
 numerical computations become impractical in the dilute regime.
It has been experimentally observed and theoretically predicted that nanotubes of radii above a critical value of about 3 nm collapse globally along their length with
a pancake-dumbell cross-section \cite{Chopra,Gao, Tang, Chang}. However, for carbon nanotubes of radii below the critical value, and therefore globally stable, we  
expect that thermal fluctuations will also cause collapse in local regions to impact their structural and transport properties.  The necking of nanowires under thermal fluctuations has similar consequences. 
DNA filaments at physiological temperature  are known to locally open to form the so called DNA bubbles~\cite{Gueron,Frank}, relevant to transcription and dentauration processes.  The size of these bubbles has been investigated by computationally intensive studies based on the Peyrard-Bishop-Dauxois model (PBD) ~\cite{Bishop, Dauxois} at
non-zero temperature~\cite{Alexandrov}, however, a theoretical understanding of their behavior is still lacking. 

We present here a very general statistical mechanics framework to study such thermally activated events which is applicable  to a variety of Q1D systems, and we show how to implement it practically to gain some insight on the phenomena described above.  We predict 
  an  exponential behavior for bubbles in DNA  (which has recently emerged from numerical model simulations but has
so far not been accessible experimentally) and the dependence of the average distance between
collapsed regions in a  nanotube or necking in a nanowire, which effectively infinite below a critical temperature and monotonically decreases to a non-zero minimum at very high temperatures. 

\begin{figure}[b!]
\hspace{1 mm} \includegraphics[width=3 in]{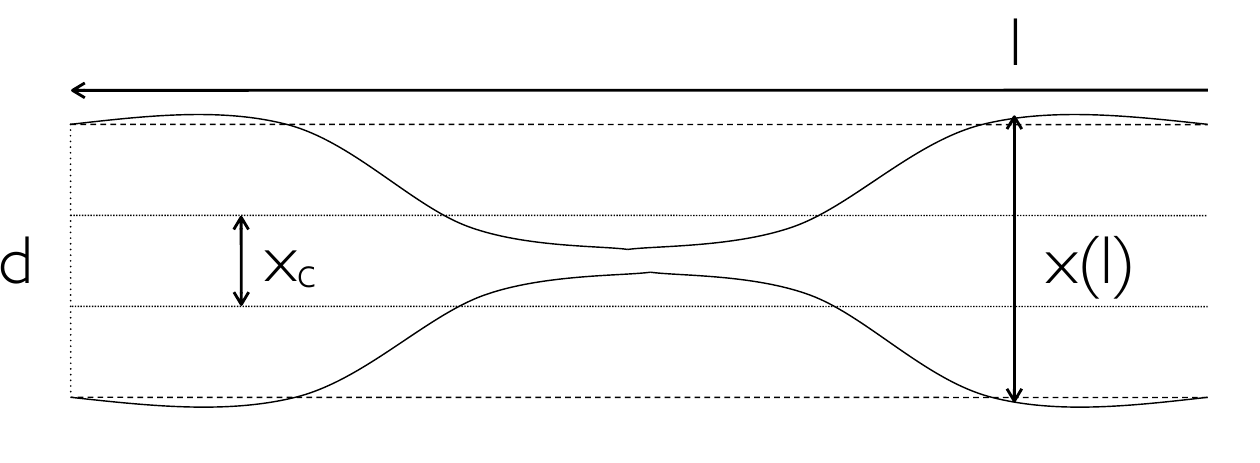}
\caption{A schematic depiction of localized failure in a Q1D system, where $x_c$ defines the collapse width, $d$ is the lateral size at equilibrium, and  $x(l)$ is the relevant degree of freedom for our problem, allowed to fluctuate in the coordinate $l$, $0\le l \le L$.}
\label{Fig0}
\end{figure}

Following previous work~\cite{Nisoli}, we consider a Q1D system like the one depicted in Fig.~\ref{Fig0}, and, for simplicity, restrict ourselves to one degree of freedom, $x(l)$, where $l$ is the coordinate along the length of the wire. The boundaries are allowed to oscillate under an elastic energy $\frac{ k }{2}x'^2(l)$ and they interact via a local potential $V(x)$.  For instance, for a collapsing  carbon nanotube, $x(l)$ represents the shorter axis of the collapse, as we described the collapsed posrtion as a ``pancake'' of heigh $x$.  In models for solitons in DNA, e.g. the celebrated Peyrard-Bishop model~\cite{Bishop}, $x$ represents the distance between complementary bases, whereas $V(x)$ is a Morse potential, chosen because exactly solvable, although a simple square well would suffice~\cite{Dauxois2}. In strained, 2-dimensional nano-islands, $V(x)$ represents the elastic interaction between the edges. 
The locality of the potential is often an approximation which is valid when the typical length occupied by  the collapse failure or instability  is longer than the lateral size of the wire. The problems we describe thus posses a characteristic  length, $l_c$ below which collapse cannot happen, simply because there is not enough available space. In a carbon nanotube $l_c$ is the length needed to go from a cylindrical to a collapsed configuration and back. Different is the case of DNA: we will be interested in studying the length of bubbles, and thus we will consider ``defect'' the joined double filament configuration: because of that    $l_c=3.3$~\AA,  the distance between bases, but there are no locality issues at play, and the potential is genuinely local,  the chemical bond between complementary bases. In nanowires, $x$ reppresent their lateral size.

This approach provides us with  a natural discretization length for our problem: that, as we shall  see later, considerably  simplifies matters.  For specificity we define ``failure'' as $x$ dropping below some critical value $x_c$, a value that is  suggested by the problem itself  and might have nothing to do with the particular shape of $V(x)$. For a nanowire it will be the threshold below which band gaps open and conductivity drops. For a carbon nanotube, it might be the distance corresponding to the metastable flattened state.
For bubble in DNA it will be the distance at which bases are linked by hydrogen bonds. 

With these assumptions, the partition function for a piece of the wire of lenght $n l_c$, starting and ending in a failure  is given by a path integral over the allowed fluctuations
\begin{equation}
Z_n=\int
{\cal D}x(l) ~e^{-\beta \int_0^{n l_c} \mathrm{d}l\left[\frac{ k }{2}x'^2(l)+  V\left(x\left(l\right)\right)\right]}
\end{equation}
with $x(0)=x(n \!\ l_c) \le x_c$, and $n$ integer. The normalization in ${\cal D} x(t)$ is chosen so that  standard transformations~\cite{Kleinert, Feynman} allow us to write $Z_n$ as density matrix
\begin{equation} 
Z_n=x_c^{-2}\int_0^{x_c}  \mathrm{d}x  \int_0^{x_c} \mathrm{d}y \!\   \langle x \mid e^{-n   \hat{H}_{\beta}} \mid y \rangle 
\label{trace}
\end{equation}
of the one-particle, temperature dependent hamiltonian given by 
\begin{equation}
\hat{H}_{\beta}=-\frac{l_c}{2 \beta k }\frac{\mathrm{d^2}}{\mathrm{d}x^2 }+\beta l_c V(x).
\label{schrodinger}
\end{equation}
Now, the probability that a system of length $n l_c$ will not fail anywhere in the middle is given by
\begin{equation}
P_n=\frac{\Omega_n}{Z_n}
\label{pro}
\end{equation}
where $\Omega_n$ is the sum over fluctuations such that $x(l)>x_c$: that is, the sum over the fluctuations that  only lead to failure at the boundaries, or
\begin{equation}
\Omega_n=\int_{x(l)>x_c}
{\cal D}x(l) ~e^{-\beta \int_0^{n l_c} \mathrm{d}l\left[\frac{ k }{2}x'^2(l)+  V\left(x\left(l\right)\right)\right]} ~,
\end{equation}
again with  $x(0)=x(n l_c)\le x_c$.
 
We divide all the fluctuations in the definition of $Z_n$ into $n$ groups: those that never lead to failure, those that lead to failure after the first $n-1$, $n-2$,  \dots,  2, 1 steps. In this way $\Omega_n$ enters the convolution equation 
\begin{equation}
Z_n=\Omega_n+v \sum_{m=1}^{n-1}\Omega_m Z_{n-m},
\label{Dyson}
\end{equation}
reminiscent of the work of M. E. Fisher {\it et al.}~\cite{Fisher}, 
The vertex term $v$, a  length, is introduced to account for the 
path integration between 0 and $x_c$  around each $m\!\ l_c$. For small $x_c$ one expects $v=\lambda x_c$. 


We can solve~(\ref{Dyson}) for $\Omega_n$ by  deconvoluting  via the generating functions $\tilde Z(u)=\sum_{n=1}^{\infty} Z_n u^n$, $ \Omega(u)=\sum_{n=1}^{\infty} \Omega_n u^n$
thus obtaining  
\begin{equation}
\tilde \Omega(u)=\frac{\tilde Z(u)}{1+v \tilde Z(u)},
\label{Oz}
\end{equation}
from which $\Omega_n$ can be derived by power expansion of  $\tilde \Omega(u)$. For instance, by direct differentiation of (\ref{Oz}), and using $\tilde Z(0)=0$ one immediately finds 
\begin{equation}
P_1=1.
\end{equation}
As expected by definition, a piece of length $l_c$ cannot fail.

For simplicity, let us assume (justifications follow later) that only the temperature dependent effective Hamiltonian in equation~(\ref{schrodinger}) has a bound ground state $\psi_0(x)$ of energy $\epsilon_o$ (functions of temperature) and that $Z_n$ reduces to a projector on that ground state, Then we can write $Z_n=(1-p)  u_0^{-n} v^{-1} 
$
where
\begin{equation}
1-p=v x_c^{-2} \Big\arrowvert \int_0^{x_c}  \psi_0(x)   \mathrm{d}x \Big\arrowvert^2,
\label{p}
\end{equation}
and  $u_0=\exp( \epsilon_0)$. Summing simple series and following the procedures described above one finds 
$\Omega_n=v^{-1}(1-p)p^{n-1}u_0^{-n}$ and 
 immediately 
\begin{equation}
P_n=p^{n-1}
\label{exp}
\end{equation}
where, correctly, $P_1=1$ and thus again, a piece of length $l_c$ does not fail. This exponential behavior of (\ref{exp}) has been found in numerical studies of bubbles in DNA, as explained later~\cite{Alexandrov}. From the average $\bar n$ one finds the  average distance $\bar l = l_c \bar n$ between failures for a long wire 
\begin{equation}
\bar l =\frac{l_c}{1-p}
\end{equation}
which shows that the lower is $1-p$ (related to the probability of penetration below $x_c$ of the eigenstate) the longer is the distance between failures. Also  the probability  $P_n$, and hence $\bar l$ or $\bar n$ are independent of $u_0$ and thus of the actual eigenvalue of the state, as expected, since the problem is invariant under uniform shift of the spectrum of (\ref{schrodinger}): of course when more than one eigenvalue is relevant,  then only the differences among them  plays a role.  

When can we apply the outmost bound eigenvalue approximation? We leave to the reader to prove that from (\ref{Oz}), when the entire spectrum of (\ref{schrodinger}) is retained, $\tilde \Omega(u)$ is a meromorphic function with real positive poles of order one. In the approximation of infrequent failures, or large bubbles in DNA (acceptable at physiological temperature)  where the average distance $\bar n$ is large, only the smallest pole is relevant. Then, if the spectrum of (\ref{schrodinger}) is such that $ \epsilon_s-\epsilon_0\gg1$, one only considers the bound groundstate, as the correction to $p$ is proportional to $\exp(\epsilon_0-\epsilon_1)$. We will see that in the cases we study both approximations are fulfilled.

\begin{figure}[t!]
\hspace{8 mm}\includegraphics[width=2.45 in]{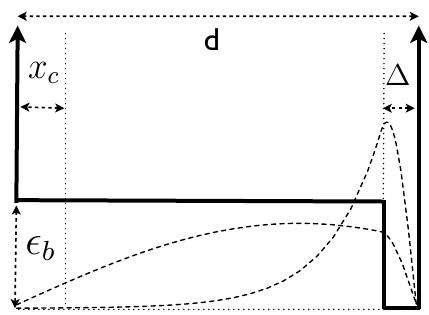}\vspace{4 mm}
\includegraphics[width=2.7 in]{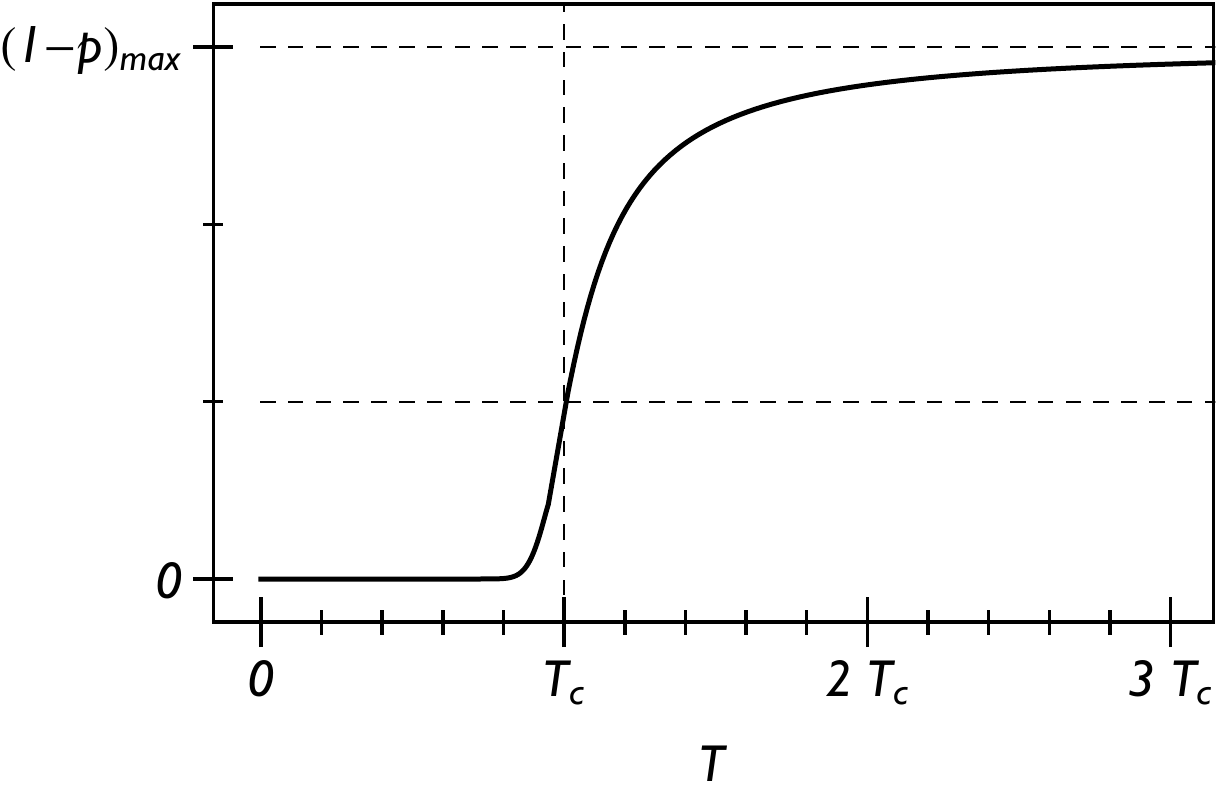}
\caption{Top: Plot of $V(x)$ in (\ref{schrodinger}). The system is in equilibrium between $d$ and $d-\Delta$, with a hard core repulsion at $d$ and zero, and a square potential barrier at $d-\Delta$ of strength $\epsilon_o$. 
Dashed lines are eigenstate of eigenvalue larger and smaller than $\beta \epsilon_b$. Bottom: the behavior of $1-p(T)\propto\bar l^{-1}$, for the potential above: notice saturation at high temperature, whereas defects are virtually inexistent at for $T/T_c< .8$.}
\label{V}
\end{figure}

For specificity we consider now the steplike potential depicted in Fig.~\ref{V}, designed to capture in a simplified picture the essential 
features of carbon nanotube collapse and  the rich phenomenology of nanowire necking   (whereas  a different approach will be described later for DNA bubbles). We assume  that the system is in equilibrium between  distances $d, d-\Delta$, with $\Delta \ll d$ as the inaccessible region is much wider than the size of the Q1D system. The infinite repulsion at $d$ ensures stability of the wire toward large fluctuations, and, in the case of carbon nanotubes, it accounts for the very large energy needed to break sp$^2$ bonds, whereas the finite height $\epsilon_b$ of the square repulsion allows for some penetration and thus a non zero probability of localized failure $1-p$ in (\ref{p}). Hard core repulsion at zero prevents negative values of $x$, which would lead to an artificial additional extensive entropy. 

Now, a look at~(\ref{schrodinger}) shows that for very small $T$, the potential barrier $\beta \epsilon_b$ is very large in the Schr\"odinger operator, and thus the probability $1-p$ from (\ref{p}) is very small, because the ground state provides little tunneling. As the temperature increases, the barrier decreases, the ground state has larger probability for penetration  and thus $1-p$ also increases. There is a definite temperature $T_c$ such that when $T>T_c$, there are no more bound states of eigenvalue less than $\beta \epsilon_b$ and the behavior of the ground state in the forbidden zone becomes sinusoidal rather than exponential, and thus $1-p$ becomes much larger.
Yet $1-p$ {\it does not increase indefinitely with temperature, but  saturates} to a value easy to compute: at large $T$,  $\beta \epsilon_p$  goes to zero, the potential tends to a flat infinite well of size $d$, and thus the ground state reduces  to  $\psi(x)=\sqrt{2/d}\sin \left( x \pi/d\right)$ from which one finds immediately, for small $x_c$,  the saturation probability 
\begin{equation}
(1-p)_{\mathrm{max}}=\frac{\pi^2}{2}\frac{ v x_c^2}{d^3}.
\end{equation}
There is therefore {\it a minimum distance} between necking of nanowires or localized collapses of nanotubes, given by
\begin{equation}
\bar l_{\mathrm{min}}=\frac{2}{\pi^3} \frac{l_c d^3}{x_c^3 \lambda},
\label{lmin}
\end{equation} 
and attained at very large temperature. This is a general feature of localized failures in Q1D systems whose lateral stability is protected by an infinite potential barrier. This is  clearly not the case for DNA, where instead denaturation, or complete separation of the two filaments, is known to occur, as we will see later.  No doubt this saturation effect could disappear should the defects come close enough to interact attractively--an interaction we completely neglect. Our estimates below show that this is not the case for carbon nanotubes.

$T_c$ can be found by solving graphically for the eigenvalue equation of the potential in Fig.~\ref{V}. If $\Delta \ll d$, states with energy less than $\beta \epsilon_p$ disappear at 
\begin{equation}
T_c=\Delta \sqrt{8k \epsilon_b}/\pi.
\label{Tc}
\end{equation}
Numerical solution of the eigenvalue equation  (Fig.~\ref{V}) shows that until about $0.8~T_c$ no defects are present, yet $1-p$ grows steeply around $T_c$  [where it is roughly $(1-p(T_c))\simeq (1-p)_{\mathrm{max}}/3$].

While the phenomenology  of nanowires is too rich to be considered here, we can use~(\ref{lmin}) for reasonable estimates in the case of carbon nanotubes  collapse.  If the radius is below a certain critical value ($\simeq 3$ nm), yet not too little ($\simeq 1$ nm),  the cylindrical configuration is globally stable~\cite{Gao,Tang}, yet the nanotube can collapse in a finite region. We  first check our approximations. From (\ref{Tc}) we have that  if $\Delta$ is less then an Angstrom, and \mbox{$k\sim\epsilon_b\sim 10^{-1}$ eV \AA$^{-1}$}, then we obtain $T_c\sim10^{-2}$ eV, or of the order of $10^2$--$10^3$ K. At very low temperatures the eigenstates reduce to plane waves in an infinite well of width $\Delta$, and their eigenvalues are proportional to $T$, which is small: nevertheless, with the numbers above one finds $\epsilon_1-\epsilon_0 \gg1$ if $T \gg 1$ K, and thus the ground state approximation works at any practical temperature. Taking $x_c=0.34$ nm (the equilibrium distance in the van der Waals interaction between graphene sheets), $d=2.8$ nm $<d_c$  ($d_c$ is the critical value above which the globally collapsed configuration is stable, between 3 and 3.5 nm~\cite{Gao,Tang,Chang}) $\lambda \sim 1$, $l_c\sim 1$ nm,  one finds  the  high temperature average length of non collapsed portions to be  $\bar l_{\mathrm{min}}\sim 100$ nm, or less. A few considerations: as $d$ comes close to $d_c$, the metastability of $V(x)$ around $x_c$ becomes significant and the potential in Fig.~\ref{V} does not properly describe the situation anymore. Conversely, as $d$ gets smaller, $\bar l_{\mathrm{min}}$ also decreases, as smaller amplitude is needed in thermal fluctuations to induce failure, yet the potential barrier $\epsilon_b$ increases, and with it the temperature $T_c$. Finally a reversed case exist for large ($d>d_c$), fully collapse nanotubes, which, conversely,  we expect to show local thermally inflated regions. From (\ref{lmin}) one can see that in this case, as $d$ is larger, so is $\bar l_{\mathrm{min}}$, of the order of many hundreds of nanometers. 

Many theoretical and computational studies have been devoted to the study of bubbles in DNA~\cite{prot,Bishop,Dauxois,Gueron, Frank, Alexandrov}, because of their importance for genetic  transcription. In an inverted approach, now it is  the joined double strand that we considered a ``defect``, and we ask ourselves what is the probability of existence of a bubble, or filament separation, of a certain length. We can use the formalism above by considering a square potential ~\cite{Dauxois2}, of width $x_c=0.5$~\AA~and depth $\epsilon_b=0.33$ eV, ~\cite{Bishop}, roughly representing the chemical bonding between complementary bases,  on a semi-infinite line. As for the elastic constant we use $k=3~10^{-3}$ eV/\AA$^2$~\cite{Bishop}: that is consistent with the disappearance of a bond state and thus denaturation at a certain critical temperature $T_c=x_c \sqrt{8k \epsilon_b}/\pi$. With the numbers above, we can see that our single eigenvalue approximation is justified for any reasonable temperature; therefore we find that the probability of a bubble decays exponentially with its length as $P_n=p^n$, a result confirmed by computationally intensive numerical studies~\cite{Alexandrov}. Also, since $p$ increases when the localization of the ground state in the potential well decreases, the average length of a bubble will be larger in the region where  the bond between complementary bases is weaker, as expected for instance in regions rich of softer AT bases, a phenomenon demonstrated by recent work~\cite{Choi}. Also $\bar l$ increases with temperature, to reach the value $p=1$, thus $\bar l=\infty$ at the critical temperature for denaturation. For temperatures very close to denaturation, the single eigenvalue approximation might break down, and the probability might not then behave exponentially.

In conclusion we have calculated the statistics of thermally-induced localized structural failures, defects or instabilities in Q1D systems.  For the collapse of carbon nanotubes, and necking of nanowires, we predict a critical temperature below which the occurrence of failure becomes extremely rare.  We also show that the behavior of the average distance between defects is monotonically decreasing in temperature and approaches a minimum avearage distance for very high temperatures. These results may be verified by experimental measurements of transport at different temperatures. For DNA, we predict that  the probability of bubble openings decreases exponentially with the length of the bubble.  It increases with  temperature and is likelier to occur where  the  strength of bonding of complementary bases is weaker. These results may be verified by simulations using the PBD model.

Discussions on DNA modeling with Boian Alexandrov (Los Alamos National Laboratory) was as useful as pleasant. This work was carried out under the auspices of the National Nuclear Security Administration of the U.S. Department of Energy at Los Alamos National Laboratory under Contract No. DE-AC52-06NA25396.


\begin{thebibliography}{}
\bibitem{disclaimer} While for tubes of radius above the critical value for global collapsing the collapsed shape is believed to be a  ``dumbbell'' our ``pancake'' approximation is likely not far from the actual shape for local collapse in smaller tubes.
\end{thebibliography}
\end{document}